\documentclass[12pt,preprint]{aastex}

\begin{document}

\title{On the transition from nuclear-cluster to black-hole dominated galaxy cores}

\author{Kenji Bekki} 
\affil{
ICRAR,
M468,
The University of Western Australia
35 Stirling Highway, Crawley
Western Australia, 6009
}

\and

\author{Alister W. Graham}
\affil{
Centre for Astrophysics and Supercomputing, Swinburne University of
Technology, Hawthorn, Victoria 3122, Australia\\}

\begin{abstract}

Giant elliptical galaxies, believed to be built from the  merger of
lesser galaxies, are known to house a massive black hole at their
center rather than a compact star cluster.  If low- and
intermediate-mass galaxies do indeed partake in the hierarchical
merger scenario, then one needs to explain why their dense nuclear
star clusters are not preserved in merger events.  A valuable clue may
the recent revelation that nuclear star clusters and massive black
holes frequently {\it co-exist} in intermediate mass bulges and elliptical
galaxies.  In an effort to understand the physical mechanism
responsible for the disappearance of nuclear star clusters, we have
numerically investigated the evolution of merging star clusters with
seed black holes.  Using black holes that are  1-5\% of their host
nuclear cluster mass, we reveal how their binary coalescence during a
merger dynamically heats the newly wed star cluster, expanding it, significantly lowering
its central stellar density, and thus making it susceptible to tidal
destruction during galaxy merging.  Moreover, this mechanism provides
a pathway to explain the observed reduction in the nucleus-to-galaxy
stellar mass ratio as one proceeds from dwarf to giant elliptical
galaxies.

\end{abstract}


\section{Introduction}

The central regions of inactive galaxies are receiving increasing attention as
astronomers begin to
accurately quantify their inner-most features.  These range from partially
evacuated cores housing a massive black hole (MBH) in giant galaxies to
excess light in the form of a dense nuclear star cluster (NC)  in less massive
spheroids\footnote{The term spheroid is used to denote either an elliptical
  galaxy or  the bulge of a disk galaxy.}.  Curiously, an unexpected connection
between MBHs and NCs  is starting to
emerge (Ferrarese et al.\ 2006a; Wehner \& Harris 2006; Balcells et al.\ 2007;
Graham \& Spitler 2009).

At the low mass end, dwarf elliptical (dE) galaxies are frequently observed to
contain a dense cluster of stars near their centre (e.g.\ 
Sandage \& Binggeli 1984; Binggeli et al.\ 1985; Ferguson \& Binggeli 1994).  The stellar mass
of these NCs relative to their host spheroid's stellar mass is known to
systematically decrease as the dE mass increases (e.g.\ Graham \&
Guzm\'an 2003; Grant et al.\ 2005).\footnote{The bulges of disc galaxies also
  commonly contain a NC (e.g., Philips et al.\ 1996; Carollo, Stiavelli \& Mack
  1998; B\"oker et al.\ 2002), and the same general trend in the 
nuclear-to-spheroid stellar flux
  ratio is observed (e.g., Balcells et al.\ 2003, 2007).}  Therefore, if
nucleated elliptical galaxies are players in an hierarchical Universe (White
\& Rees 1978), one cannot simply merge such galaxies and double the mass of
the new host galaxy and its NC.

At the high mass end are massive elliptical galaxies --- 
the end product of major mergers.  However,
such galaxies are observed not to contain NCs, instead they possess central
stellar deficits relative to the inward extrapolation of their outer S\'ersic
light profile (e.g.\ Graham et al.\ 2003; Trujillo et al.\ 2004; Graham 2004;
Ferrarese et al. 2006b).
While it has been advocated that the dry merging of elliptical galaxies will result in
the partial evacuation of the new galaxy's core due to the binary coalescence
of pre-existing MBHs (Begelman et al.\ 1980; Ebisuzaki et al.\ 1991;
Milosavljevi{\'c} \&  Merritt 2001;
Merritt \& Milosavljevi{\'c}  2005; Merritt et al.\ 2007), there
must be more going on.  There must be a phase which erases the NCs 
---  not included in the above mentioned studies --- 
that are prevalent in the less massive progenitor galaxies.
While Berczik et al.\ (2005) used Plummer models to represent galaxies in
their detailed analysis of the impact that binary MBHs can have, here we dramatically
rescale the problem, using Plummer models to represent NCs that are
$\sim$$10^3$ times smaller and, globally, $\sim$$10^5$ times denser than galaxies.

Our motivation arises because 
it has recently been recognised that intermediate mass elliptical galaxies,
and the similarly massive bulges of disc galaxies, regularly contain both a NC
and massive black hole (e.g.
Graham \& Driver 2007; \ Gonz{\'a}lez Delgado et al.\ 2008; Seth et al.\
2008).  Graham \& Spitler (2009) have quantified how
the $M_{\rm BH}/M_{\rm NC}$ mass ratio ($F_{\rm BH}$)
increases with increasing host spheroid stellar
mass, $M_{\rm sph}$, 
until only a MBH is present at the centre.  While the runaway merger of
NC stars during a merger event may lead to their conversion into a
MBH (e.g., Zel'dovich \& Podurets 1965; Frank \& Rees 1976; Quinlan \& Shapiro
1987; Lee 1993), and feedback processes may also impact $F_{\rm BH}$
(McLaughlin et al.\ 2006; Nayakshin et al.\ 2009), this Letter explores whether dense NCs with seed MBHs might 
evaporate during a collision due to dynamical heating by the MBHs.

\section{The model}

Working from an established $N$-body code (Bekki et al.\ 2004; Bekki 2010)
which runs on the GRAvity PipE (Sugimoto et al.\ 1990), 
we have developed an idealized model in which a new single NC can be formed from
the collisionless merger of two NCs with MBHs --- 
an event likely to occur during a major galaxy
merger (e.g., Bekki 2007a; Bekki et al.\ 2010 in preparation). 
Here we investigate how the final structure of the new NC depends on the
mass ratio $M_{\rm BH}/M_{\rm NC}$ ($=F_{\rm BH}$) of the initial NCs. 
We assume that the dynamical evolution of the two 
NCs are dominated  by the NCs and MBHs themselves, rather than by the
gravitational field of background stars.  Thus, each of the present models
includes only two NCs and two MBHs: it includes neither background field stars
nor external tidal fields of galaxy mergers.

%
The total  mass and size  of an initial NC are represented by
$M_{\rm NC}$ and $R_{\rm NC}$, respectively.
All masses and lengths are measured in units
of $M_{\rm NC}$ and $R_{\rm NC}$ unless otherwise
specified. Velocity and time are measured in units of $v$ = $
(GM_{\rm NC}/R_{\rm NC})^{1/2}$ and $t_{\rm dyn}$ = $(R_{\rm
NC}^{3}/GM_{\rm NC})^{1/2}$, respectively, and the gravitational
constant $G$ is assumed to be 1.
If we adopt $M_{\rm NC}$ = 5.1 $\times$ $10^{6}$ $ \rm M_{\odot}$ 
and $R_{\rm NC}$ = 77 pc as fiducial values 
--- corresponding to $\omega$ Cen (e.g., Meylan et al.\ 1995),
which is considered to originate from a nucleated galaxy (e.g., Bekki \&
Freeman 2003) --- then
$v$ = 16.9 km s$^{-1}$ and $t_{\rm dyn}$ = 4.46
$\times$ $10^{6}$ yr.
The gravitational softening length, ${\epsilon}_{\rm g}$, is set equal to the
mean separation of stellar particles at the half-mass radius\footnote{The 
  half-mass radius equals $0.25R_{\rm NC}$.} of the initial NC: 
${\epsilon}_{\rm g}=0.01R_{\rm NC}$ (=0.77 pc).  This
softening length is also adopted for the MBHs.

The radial density profile of our preliminary NC is given by a Plummer model
with scale length set equal to $0.2R_{\rm NC}$. 
We will explore in detail  how the present results  depend on 
models with different initial radial profiles
(e.g., King or S\'ersic models) in future work. 
To construct a model in dynamical equilibrium  for a NC with
a MBH located  at its center, we adopt the following two steps.
First, the initial  mass of the MBH  in our isolated  NC model
is set to the mass of 
each individual star ($m_{\rm star}$) in the NC.  Second,  we run the  isolated model
such that 
the initial MBH  mass ($m_{\rm star}$)  is  increased steadily and  slowly
to finally reach any adopted $M_{\rm BH}$ value within $20$ $t_{\rm dyn}$
(i.e., adiabatic growth of the MBH).
During this isolated evolution of the NC, the  stellar distribution 
of the NC can adiabatically evolve into a new dynamical 
equilibrium. 
We then use this new radial distribution of the stars 
for our progenitor NCs which are subsequently merged.

We have confirmed that with time steps as small as 
$4.5 \times 10^4$ yr for models with $F_{\rm BH}\le 0.025$,
the NCs with MBHs are stable after $20$ $t_{\rm dyn}$. 
The changes to the central stellar densities due to adiabatic MBH
growth in models with MBHs are only a factor
of $\sim 2$ in comparison with those with no MBH. 
This effect is much smaller than the factor of ten change in central stellar 
density due to MBH heating, as shown later. 
We are therefore able to probe the effects of MBHs in NC merger
remnants on the inner stellar densities of the remnants.

The two NCs in a NC merger are referred to as NC1 and NC2 and the relative
positions and velocities of NC2 with respect to NC1 are
set to be ($X_{\rm r}$, $Y_{\rm r}$, $Z_{\rm r}$)
and ($U_{\rm r}$, $V_{\rm r}$, $W_{\rm r}$), respectively.
Although the relative positions and velocities of NC2
are free parameters, and we investigated models with different values
for these 6 parameters,
we only show the results of the models with
($X_{\rm r}$, $Y_{\rm r}$, $Z_{\rm r}$)=(4, 0.5, 0)
and 
($U_{\rm r}$, $V_{\rm r}$, $W_{\rm r}$)=(-1, 0, 0). 
The total number of stellar particles used in a  model for NC merging  is
$4 \times 10^5$, allowing us to  conduct a large parameter study
(e.g., $F_{\rm BH}$ and $Y_{\rm r}$, and $U_{\rm r}$)  for NC evolution
with MBHs.
We follow the dynamical  evolution  of two merging  NCs for $20$ $ t_{\rm dyn}$
within which the two NCs merge with each other completely
to form a new NC.

The two MBHs in the newly formed NC   can
drift around the central region of the NC after the BHs form a very
close pair owing to their orbital decay caused by dynamical friction
against the NC stars.   
We note that a Newtonian gravitational force is
always assumed (outside of ${\epsilon}_{\rm g}$)
 during simulations
with GRAPE 
(we do not investigate MBH merging through gravitational wave radiation). 
We assume 
that the MBH pair can merge to form a single MBH akin to the  merging of stellar-mass BHs in
dense star clusters (e.g., Quinlan \& Shapiro 1989) and adopt the
following two steps to obtain the final stellar distribution in the NC merger.
First, the MBHs in the merger remnant 
 are replaced with a single MBH
with position and velocity equal to the mass center of the
two MBHs. This is done after $20$ $t_{\rm dyn}$, when the single NC is already 
formed and dynamically relaxed. 
Second, we follow the evolution of the NC merger
remnant with the new MBH for a further $20$ $ t_{\rm dyn}$  of the original NCs,
so that the single MBH can sink into
the center of the remnant due to dynamical friction and  the stellar
distribution can change in response.  We use this final stellar
distribution for the investigation of the radial density profile of the NC
merger.

The above models are referred to as ``single merger models''.  We also ran
``sequential merger models'' in which radial density profiles of stars in
stellar remnants of two and three sequential NC mergers  are investigated. The
progenitor NCs of the second NC merger are the remnant of the first NC
merging, and those of the third sequential merger  are the remnant of the
second sequential NC merging, 
with one key difference being that these mergers are not evolved for 20 $t_{\rm
  dyn}$ between successive mergers. 
In order to better understand the physical
role of dynamical heating on NCs by MBHs,
 we also ran single and sequential
merger models with no MBHs (i.e., $F_{\rm BH}=0$).  

The present study can be compared with Ebisuzaki et al.\ (1991) who
investigated the radial density profiles of elliptical galaxies formed from
galaxy merging with MBHs.  Although they did not investigate the dynamical
influence of MBHs on NCs, they clearly pointed out that MBHs can lower the
inner densities of giant ellipticals because of the dynamical effects of MBHs
(Begelman et al.\ 1980).  One of the significant differences between their
work and ours is that we explore models with much larger $F_{\rm BH}$ (up to
0.05, although see Kandrup et al. 2003).

\section{Results}

Fig.\ 1 shows that the final internal (three-dimensional, 3D) radial density profiles
${\rho}_{\rm NC}(r)$ are significantly different between our four single merger
models with different $F_{\rm BH}$.  
The final central 3D densities depend on $F_{\rm BH}$ such that they are lower in
models with larger $F_{\rm BH}$.  
For example, the inner stellar density at
$r/R_{\rm NC}=0.05$ ($\log r/R_{\rm NC}= -1.3$) in the model with $F_{\rm
  BH}=0.05$ is a factor of $\sim 32$ lower than that in the model with $F_{\rm
  BH}=0$ (i.e., with no MBH). This reflects how dynamical heating of NCs, by MBHs
during NC merging, expels stars from the central
regions and consequently lowers the inner stellar densities of newly wed
NCs. A significant fraction of stars initially within $R_{\rm NC}$  can be relocated
well beyond  $R=5R_{\rm NC}$: the fraction can be as large as 0.37
in models with $F_{\rm BH}=0.05$. 
Moreover, the internal radial density profiles of merger
remnants,  for $r/R_{\rm NC} \le \sim1/4$,
are  shallower in models
with larger $F_{\rm BH}$.  These results clearly demonstrate that 
MBHs ($F_{\rm BH}$) can control the stellar structure in (collisionless) mergers of NCs
owing to dynamical heating  by MBHs.

To understand why the stellar mass densities in models with MBHs 
can be significantly smaller than those without MBHs (as shown in Fig.\ 1),
we ran comparative models with $F_{\rm BH}=0.01$
in which only NC1 had a MBH (i.e., 
$F_{\rm BH}=0$ for NC2).
Fig.\ 2 
shows that 
the merger model with a MBH only for NC1 has a significantly higher 
central density 
than that in the model with MBHs in both NC1 and NC2.
This strongly suggests that dynamical heating from binary MBHs 
sinking into the merger remnant of the two NCs is important
in reducing the inner density of the remnant. We also observe 
that the merger model with a MBH only for NC1 has a smaller
inner stellar density in comparison with the model with no MBH, 
revealing that a single MBH 
can also heat up the remnant to some extent as it sinks to the center.

Fig.\ 3 illustrates that the final stellar profile of the three sequential major
merger events with $F_{\rm BH}=0.025$ has a significantly lower
central density than when $F_{\rm BH}=0$ (i.e., no MBH; compare Figure~1 by
Makino \& Ebisuzaki 1996). 
Fig.\ 3 also reveals that
the inner stellar densities of NC merger remnants in the sequential model with
$F_{\rm BH}=0.025$ can become progressively lower as NC (and possibly
inspiraling globular cluster)  
merging proceeds: the central stellar density
in the  final merger remnant (i.e., after three sequential merger events)
is  a factor of $\sim$ $16$ smaller than  in the original NC for
this sequence.

 Fig.\ 4 shows that ${\rho}_{\rm NC}$ at $r=0.05R_{\rm NC}$ is lower in the
 sequential merger models with larger $F_{\rm BH}$ owing to stronger dynamical
 heating.  We do however note that
 the model with $F_{\rm BH}=0.01$ does not show
 a  significant decrease of its stellar densities at $r=0.2R_{\rm NC}$ 
 owing to the much less effective dynamical effects of the MBHs on the NCs.
 This is in accord with the binary MBH scouring of the core regions in massive elliptical
 galaxies, in which the loss cone typically dominates only the inner few percent of the
 stellar distribution (e.g., Trujillo et al. 2004).
 These simulations also 
 suggest that larger, more massive NCs 
will be  more difficult to observe as distinct NCs as their 
 stellar densities may become comparable or less than that of their host galaxy. 
 Furthermore, Fig.\ 4 shows that ${\rho}_{\rm NC}$ at
 $r=0.2R_{\rm NC}$ (corresponding to the scale-radii  of
 the original NCs) is lower in the sequential merger models with larger $F_{\rm
  BH}$, though the dependence is weaker than that of ${\rho}_{\rm NC}$ at
 $r=0.05R_{\rm NC}$ on $F_{\rm BH}$.  This result implies that more massive NC 
merger remnants with lower central densities
 are more susceptible to tidal destruction by the external gravitational fields
 of their host galaxies owing to their lower mean stellar densities.

\section{Discussion and conclusions}

So far we  have  focused on the internal density profiles
of NCs and have not  discussed their  {\it projected} (two-dimensional, 2D)
radial density profiles,  which can be directly compared with
recent observational studies for 
(i) the origin of the apparent MBH-NC connection  (e.g., C\^ot\'e et al. 2006)
and 
(ii) the observed $F_{\rm BH}-M_{\rm sph}$ relation (Graham \& Spitler 2009).
Fig.\ 5 reveals that 
the {\it projected} radial density 
profiles of NC merger remnants, 
${\Sigma}_{\rm NC}(R)$, 
are significantly different between our four models
with different $F_{\rm BH}$.
The rather low central ${\Sigma}_{\rm NC}$ value at $R=0.05R_{\rm NC}$ 
and shallow inner density profile, 
in the model with  $F_{\rm BH}=0.05$ suggests that if this merger remnant
is located in the central region of a galaxy, it is less likely to be observed
as a distinct NC.
While the order of magnitude drop in surface density may seem like overkill,
especially given the apparently small levels of excess nuclear light seen in
most resolution-limited images, we note that well-resolved galaxies can have NC
light up to 5 mag arcsec$^{-2}$ (100$\times$) brighter than the underlying galaxy (e.g.,
Graham \& Spitler 2009). 

The present study confirms that more evolved  NCs --- by which we mean NCs, with
MBHs, that are further along the merger tree --- can 
have lower inner  densities (${\rho}_{\rm NC}$ and ${\Sigma}_{\rm NC}$) 
and shallower inner 2D density  profiles than their progenitors.  This  suggests   that
boundaries between distinct stellar nuclei and background field stars
in galaxies are less clear for more evolved  systems as the NCs are effectively
washed-out and dissolve into the host galaxy. 
Such diffuse NCs are also more susceptible to tidal destruction during galaxy
merging. 
The present study therefore suggests that the observed $f_{\rm
 BH}$-$M_{\rm sph}$ relation can be understood in terms of the structural
evolution of merging NCs with MBHs.

Measurements of partially-depleted
galaxy cores, relative to a galaxy's outer light-profile,  have revealed a
correlation between the central stellar mass deficit and the luminosity of the host
spheroid and its MBH mass (e.g., Graham 2004; Ferrarese et al.\ 2006b).  
As
detailed in Graham \& Guzm\'an (2003) and C\^ot\'e et al.\ (2007), the
transition between massive galaxies with  partially-depleted cores
and those without ---
which frequently have excess nuclear light instead --- occurs around $M_B =
-20.5$ mag.  Previous numerical simulations proposed that
the origin of these central stellar deficits can be understood
in the context of core formation through dynamical heating of stars
by inspiralling MBHs in galaxy merging (e.g., Ebisuzaki et al. 1991).
The present study has, for the first time, addressed one of the
over-looked problems related to the nuclear structures  of galaxies:
why and how can dense NCs  disappear during galaxy growth through galaxy merging ?

We advocate here that core-depletion due to the gravitational slingshot of
host galaxy stars by inspiralling MBHs will not occur in earnest until the NCs surrounding
the MBHs have first been eroded away by this same mechanism:
once  the NCs are effectively gone, the binary MBHs, perhaps from additional
 merger events, can then commence to eat into the inner light
profile of the host galaxy to produce the observed partially-depleted cores. 
This important
step can explain why NCs disappear along the spheroid mass sequence and it also
offers a process through which to understand the observed $F_{\rm BH}$-$M_{\rm
bulge}$ relationship in terms of galaxy formation
within the hierarchical merging scenario.

The present study suggests that
if  NCs in low-mass galaxies  have seed MBHs,
then their inner densities should progressively decline 
as galaxies grow through merging. 
This is at odds with the simple superposition of the NC density field for
merging NCs without MBHs (Fig.\ 3, lower panel). 
Although many previous theoretical studies investigated how 
NCs are formed, 
either by merging of SCs 
(e.g.,  Tremaine et al. 1975;  Capuzzo-Dolcetta \& Miocchi 2008)
or by dissipative gas dynamics in galaxies
(e.g., Bekki et al. 2006; Bekki 2007b),
they did not predict  (i) how BHs can be formed in NCs 
and (ii) what a reasonable
value is for $F_{\rm BH}$.
The formation of seed MBHs {\it within  NCs}
may be different from that of intermediate-mass
BHs in {\it isolated} globular clusters through merging of stellar-mass black
holes (e.g., O'Leary et al. 2006),
because the deeper gravitational potential wells of the 
NC host galaxies  would  play a role in retaining interstellar gas more
efficiently. 
It is thus our future study to investigate how seed MBHs
can be formed in NCs at the epoch
of NC formation in low-mass galaxies based on more sophisticated numerical
simulations.

\acknowledgments
We are grateful to the anonymous referee for valuable comments 
which improved the presentation of this paper.
K.B. acknowledges the financial support of the Australian Research
Council throughout the course of this work.
Numerical computations reported here were carried out both on the GRAPE
system
at the University of New South Wales  and on those
kindly made available by the Center for computational
astrophysics
(CfCA) of the National Astronomical Observatory of Japan.
This work was financially supported by CfCA.


\newpage

\begin{figure}
\epsscale{1.0}
\plotone{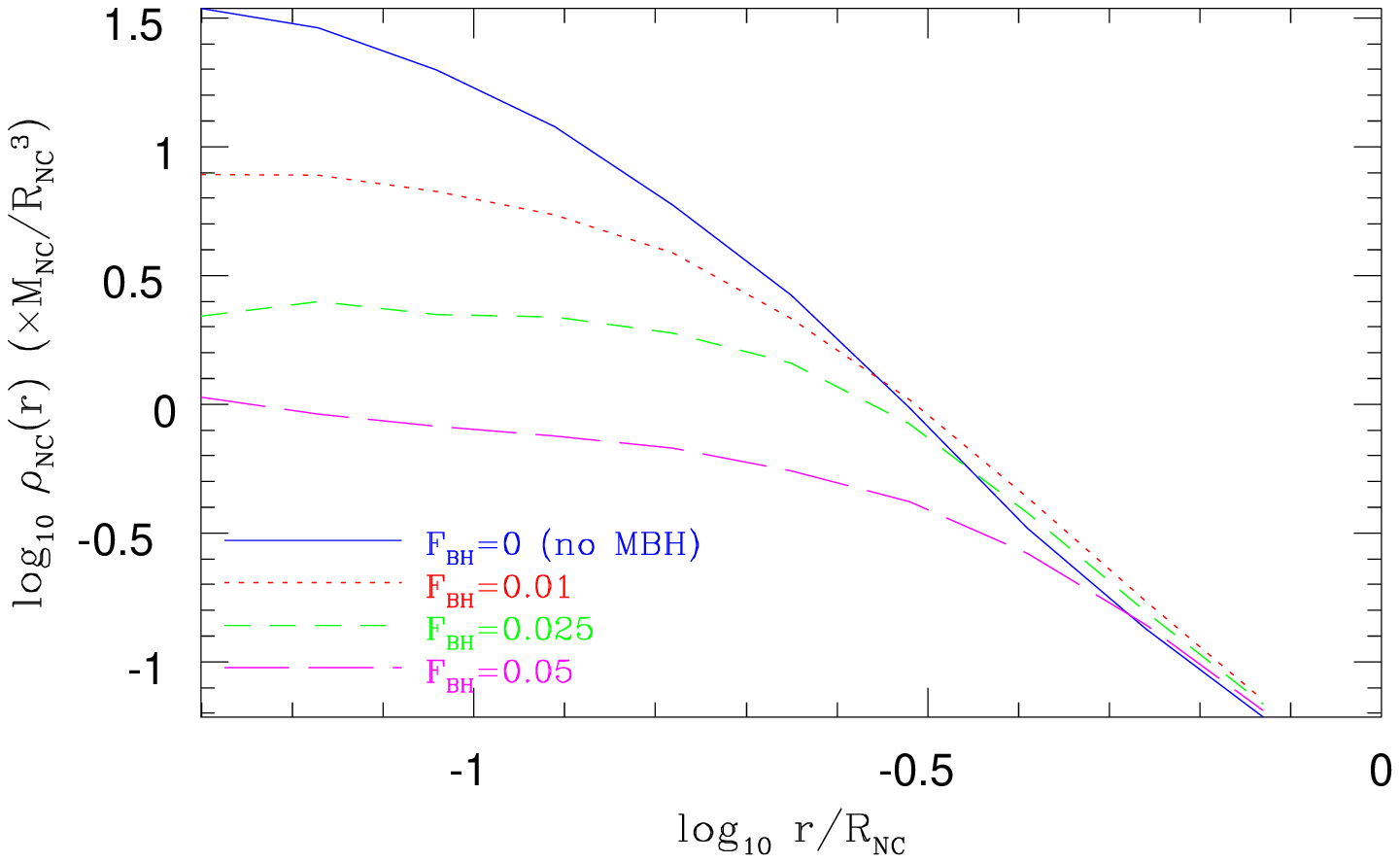}
\figcaption{
Final radial density profiles of NCs, ${\rho}_{\rm NC}(r)$, 
at an epoch $20 t_{\rm dyn}$ after MBH coalescence  
for four different single merger models.
Note that the inner densities become smaller for NC mergers 
with higher mass fractions of MBHs (i.e., larger $F_{\rm MBH}$).
\label{fig-1}}
\end{figure}

\newpage

\begin{figure}

\epsscale{1.0}
\plotone{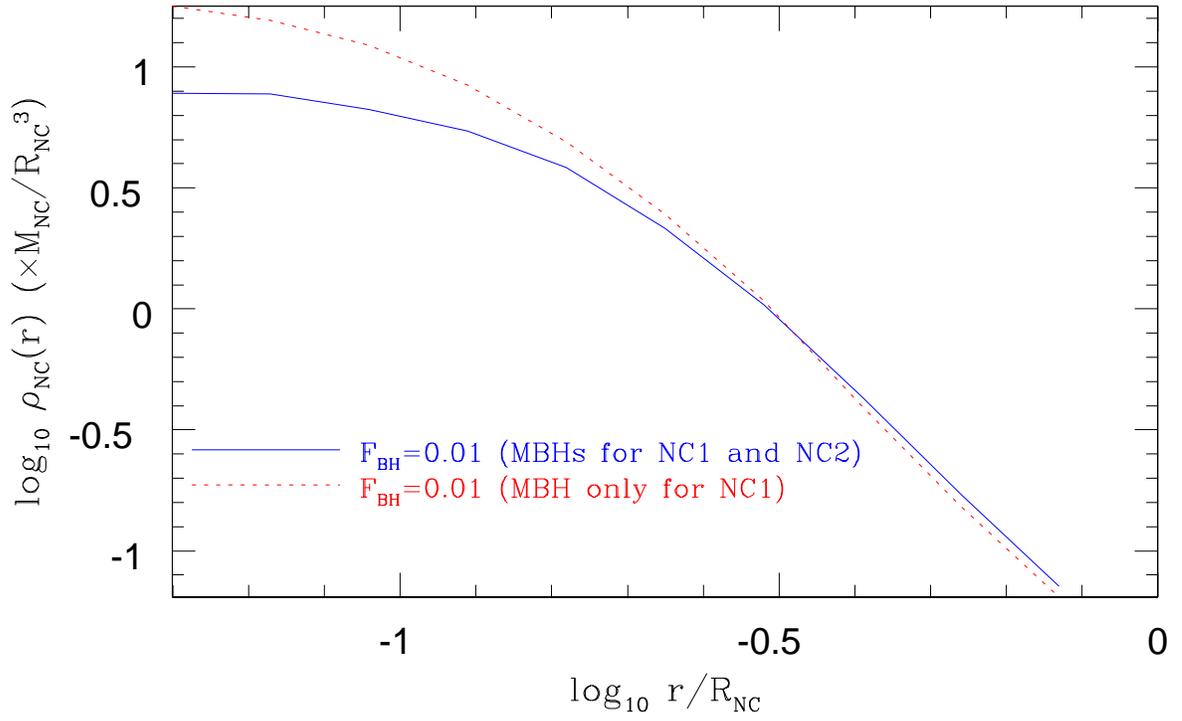}
\figcaption{
Same as Fig.\ 1 but for models with a MBH only for NC1 (red dotted)
and with MBHs both for NC1 and NC2 (blue solid).  For these models,
$F_{\rm BH}=0.01$ is used. 
\label{fig-2}}
\end{figure}

\newpage

\begin{figure}
\epsscale{1.0}
\plotone{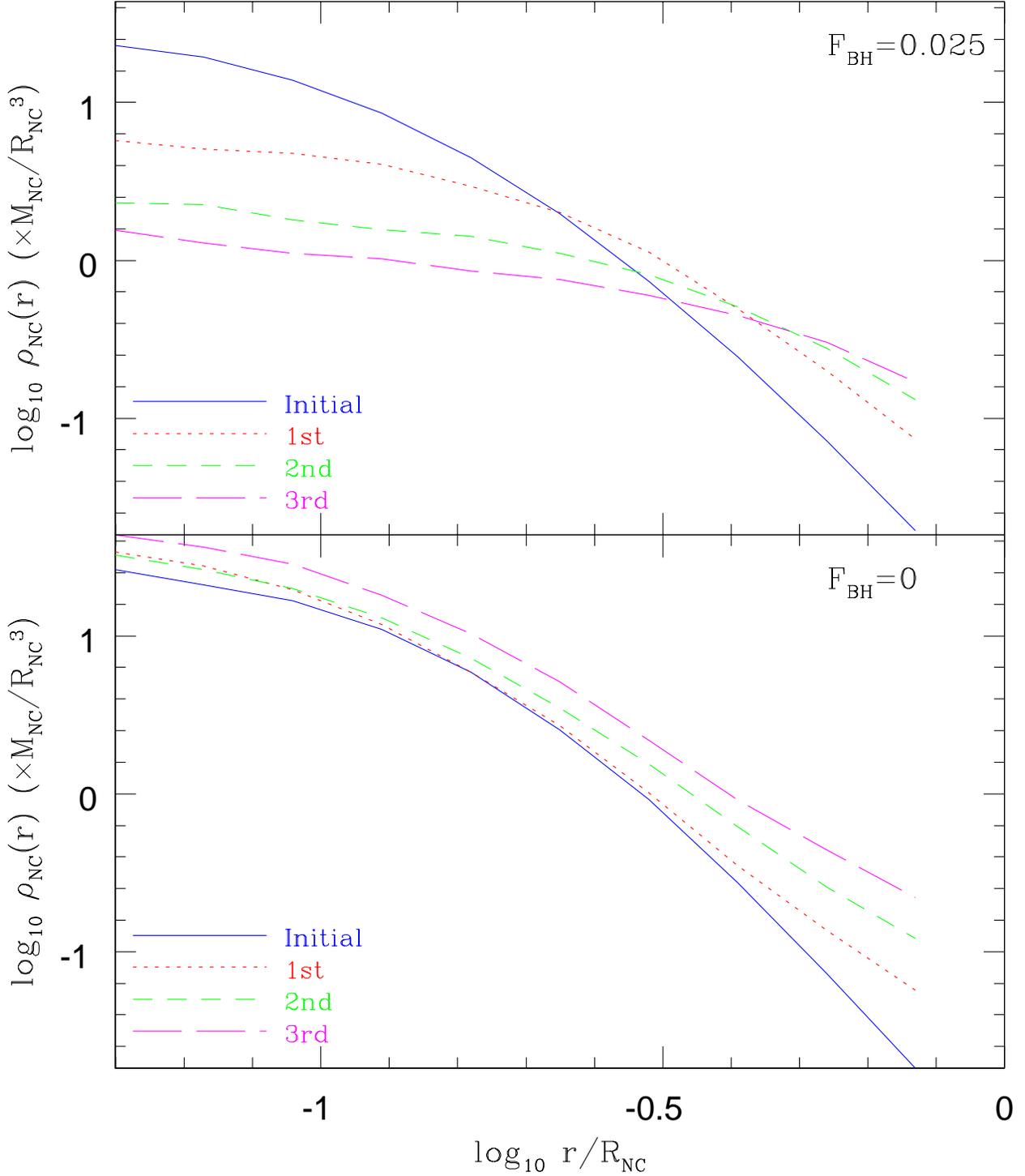}
\figcaption{
The same as  Fig.\ 1 but for two sets of sequential models.
The radial stellar distributions for the initial NCs,
the 1st merger remnant,  the 2nd, and the 3rd
are shown by the blue solid, red dotted, green short-dashed, and 
magenta long-dashed lines, respectively.
The initial profiles are slightly different between the two
sequential models because the initial radial profile
in the  model with $F_{\rm BH}=0.025$ 
is the profile $20 t_{\rm dyn}$ after adiabatic growth of the single MBH.
The dotted line in the upper panel can  be different from the short-dashed
line in Fig.\ 2 for $F_{\rm BH}=0.025$,
because the results of the model in this figure are just after merging
of NCs. 
\label{fig-3}}
\end{figure}

\begin{figure}
\epsscale{1.0}
\plotone{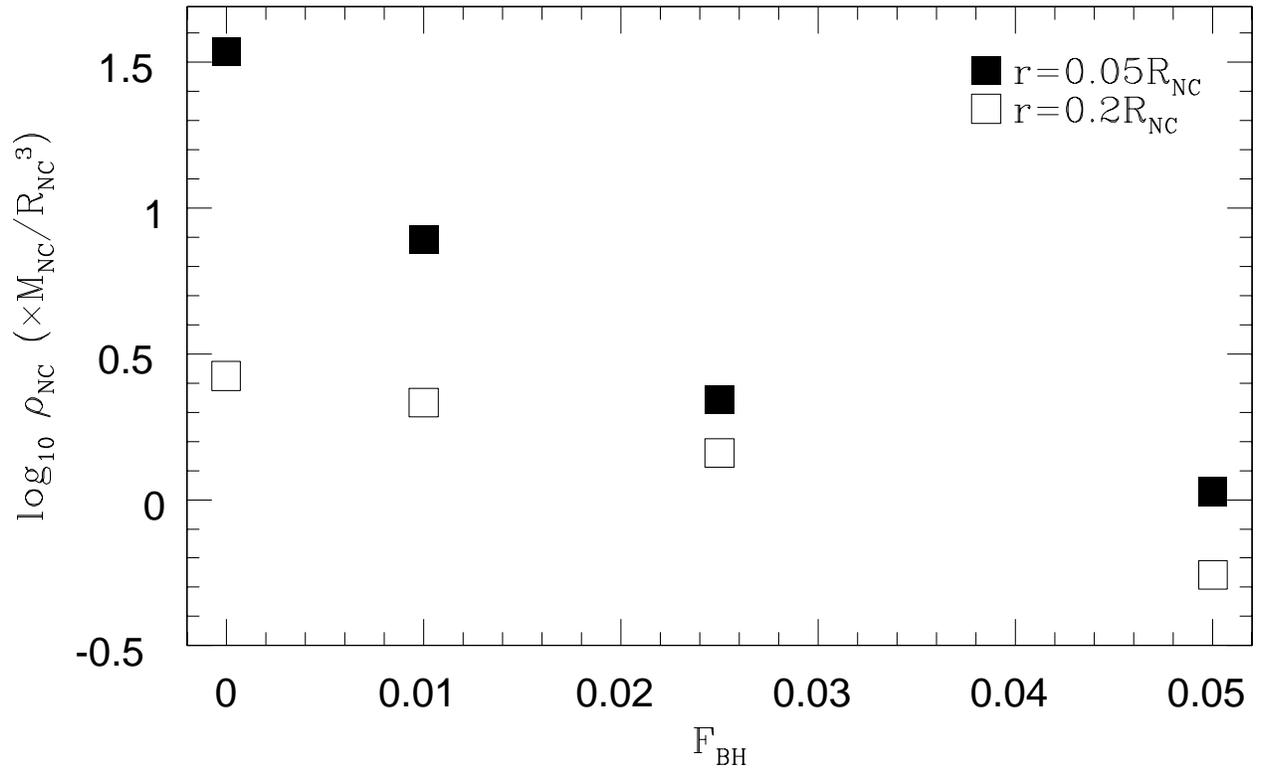}
\figcaption{
The dependence of final stellar densities 
(i.e., those from the 3rd merger remnants) at $r=0.05 R_{\rm NC}$
(filled squares) and at $r=0.2 R_{\rm NC}$ (open squares)
on the initial  $F_{\rm BH}$ mass ratio. 
\label{fig-4}}
\end{figure}

\begin{figure}
\epsscale{1.0}
\plotone{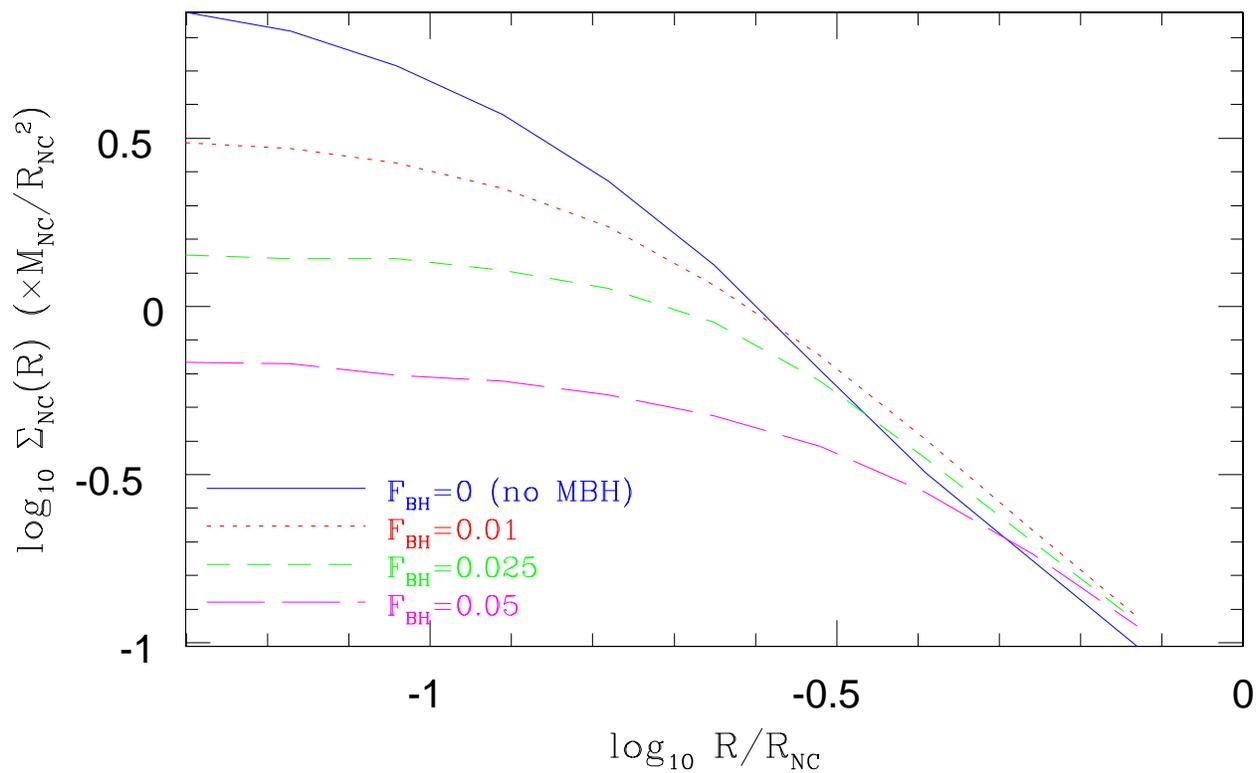}
\figcaption{
The projected radial density profiles, ${\Sigma}_{\rm NC}(R)$, 
for four different single merger models with 
$F_{\rm BH}=0$ (blue solid, i.e., no MBH),
$F_{\rm BH}=0.01$ (red dotted),
$F_{\rm BH}=0.025$ (green short-dashed),
and $F_{\rm BH}=0.05$ (magenta long-dashed).
\label{fig-5}}
\end{figure}

\end{document}